\documentclass[10pt,preprint2]{aastex}
\include{graphicx}

\shorttitle{Faults in solid neutron star matter}
\shortauthors{Jones}

\begin{document}


\title{Nature of fault planes in solid neutron star matter}


\author{P. B. Jones}
\affil{University of Oxford, Department of Physics, Denys Wilkinson
Building, Keble Road, Oxford OX1 3RH, UK}
\email{p.jones1@physics.ox.ac.uk}



\begin{abstract}
The properties of tectonic earthquake sources are compared with those
deduced here for fault planes in solid neutron-star matter.  The
conclusion that neutron-star matter cannot exhibit brittle fracture at
any temperature or magnetic field strength is significant for current
theories of pulsar glitches, and of the anomalous X-ray pulsars and
soft-gamma repeaters.
\end{abstract}




\section{Introduction}

It is widely assumed that brittle fractures caused by Maxwell or
other stresses in neutron star crusts are involved in a number
of phenomena, for example, the soft gamma repeaters (SGR), the
persistent emission of the anomalous X-ray pulsars (AXP) (Thompson et 
al 2000),
and large pulsar glitches (Ruderman, Zhu \& Chen 1998).  In current
theories of the AXP and SGR sources,
brittle fractures, propagating with a velocity of the order of
the shear-wave
velocity $c_{s}$, generate shear waves which in turn couple with
magnetospheric Alfv\'{e}n modes.  At angular frequencies
 $\omega \approx 10^{4-5}$
rad s$^{-1}$, the coupling is thought to be an efficient mechanism for
energy transfer to the magnetosphere, as shown by Blaes et al (1989).
Statistical comparisons
of SGR burst properties with those of terrestrial earthquakes
are not inconsistent with the brittle fracture assumption 
(Hurley et al 1994, Cheng et al 1996, Gogus et al 1999).  However, the
purpose of this letter is to note that elementary deductions of the
properties of neutron-star fault planes show that brittle fracture is not
possible.

\section{Neutron-star fault planes}

The relation between stability and stratification for neutron star matter
(Reisenegger \& Goldreich 1992; see also Jones 2002) constrains the
movement of matter, bounded by any fault plane which may be formed, to
an almost exactly
spherical equipotential surface.  Shear strain (for in-plane shear) is
shown in Figure 1 by the displacement of
a series of constant surfaces in Lagrangian
coordinates which intersect
a local element of the fault plane $xz$.  In the neutron star case, it might
be caused by a changing Maxwell stress reflecting evolution of the internal
magnetic flux density distribution. The components $\sigma_{ij}$ of the
stress tensor act on the surfaces of the volume elements shown on opposite
sides of the fault-plane. (We assume an isotropic elastic medium in which,
before fracture, stress propagates perfectly across the plane.)
In the brittle fracture of a terrestrial earthquake, the
stress falls at the instant of failure from $f = \sigma_{xy}$ to
a much smaller value (zero in the ideal case).  Stress energy is largely
converted to kinetic energy so that strain relaxation $\Delta\epsilon$
occurs with an acceleration $\dot{v}$ such that shear waves are
generated efficiently (see Kostrov \& Das 1988 concerning the definition
of a tectonic earthquake source).  The mechanical properties of 
terrestrial matter, as a function of depth, have been tabulated, for
example, in Kaye \& Laby (1986).  At a typical earthquake focus depth
of 15 km, the pressure $P \approx 10^{-2}\mu$, where $\mu$ is the shear
modulus.  Crack propagation is possible because $P$ is not so large
that it inhibits void formation behind the tip. At greater depths, the
increases in $T/T_{m}$, where $T_{m}$ is the melting temperature,
and in $P/\mu$ both inhibit the void formation which is necessary for
crack propagation.
Thus a transition to plastic stress-response occurs at approximately
20 km depth because dislocation glide is not pressure-inhibited
(see Scholz 1990).  Tectonic
earthquake sources are not found in the plastic region.  The deep-focus
sources, which occur in subduction zones at depths up to 500 km, are
probably intermittently running polymorphic phase transitions, following
the original idea of P. W. Bridgman (1945).

The contrast with neutron-star matter is extreme because the latter
is not absolutely stable, being in equilibrium only at finite pressure. 
Those isotropic components of the stress tensor derived from the
electrons and from Coulomb interaction (the Coulomb-electron partial
pressure $P_{Ce}$ which excludes the neutron partial pressure)
are one or two orders of magnitude
larger than the shear modulus and this largely determines the
structure of defects such as monovacancies (Jones 1999).  The way in
which the nearest neighbours to a monovacancy site relax by displacement
considerably reduces the monovacancy formation enthalpy, for example,
to 13 Mev at a matter density of $8 \times 10^{13}$ g cm$^{-3}$ in the
neutron-drip region.  Such relaxation is not possible for the nuclei
surrounding a void
whose linear dimension is of the order of, or greater than, the
relativistic electron screening length.  In any solid for which the
inequality $P_{Ce} \gg \mu$ is satisfied, it is anticipated that a void, 
even if formed, would be unstable
against dissociation to monovacancies with a lifetime, in neutron-star
matter, perhaps of the order of $10^{-21}$ s. Thus we can assert that
nuclei at a fault
plane may have different, probably less, small-distance order than
exists elsewhere but
almost the same nearest-neighbour separations.  Reference to the theory
of crack stability (Anderson 1995, see also Landau \& Lifshitz 1970)
shows that, in brittle
fracture, the important properties of the stress distribution of an
unstable crack are the
zero on the  fault plane and the singularity at its tip (in the ideal case)
which enable it
to propagate into regions of much lower stress. These do not exist
in neutron stars owing to the impossibility of forming a sufficiently
long-lived void.
 
Previous formation enthalpy calculations (Jones 2001) have shown that
an amorphous heterogeneous solid phase is formed in the crust and is
likely to persist as the star cools.  The mean square deviation
$\overline{(\Delta Z)^{2}}$ from the average nuclear charge is large compared
with unity.  The system is analogous with a multi-component amorphous
alloy in having some local order, which vanishes with increasing
length scales, but differs in having no
absolute stability.  Shear stress acting on such a system produces
faults in the form of localized shear bands.  These are
thin layers of inhomogeneuos plastic flow, formed sequentially, which
permit local strain relaxation through through dissipation and the
transfer of stress energy to neighbouring regions. 
A shear band has structure similar to a viscous
layer being deformed adiabatically between parallel plates
(see, for example, Xing et al
1999, who also give a typical stress-strain relation).
The process is not analogous with brittle fracture because
most of the local stress energy, instead of being converted to kinetic
energy, is either dissipated or transferred to neighbouring regions.
The rotation of a circular cylinder of solid crust under Maxwell stress,
by plastic flow on the cylindrical fault surface delimiting it, is an
elementary example satisfying the stratification and stability
constraint (see also Thompson et al 2000). The plastic flow threshold
is $f_{o}$ and the Maxwell stress is assumed to be
$f\exp(-s/s_{o})$, decreasing with slip distance $s$, where $s_{o}$
and $f$ are constants. 
Suppose that an unspecified stress-transfer incident results in
$\Delta f = f - f_{o} > 0$ at some instant.  The equation of motion is
\begin{eqnarray}
\frac{1}{4}\rho a \ddot{s} = \Delta f - f\frac{s}{s_{o}}
\end{eqnarray}
where $\rho$ is the matter density and $a$ the cylinder radius.
The velocity at time $t$ during slip through a distance
$2s_{p}$, where $s_{p}=s_{o}\Delta f/f$, is
$v=\dot{s}=\tilde{\omega}s_{p}\sin(\tilde{\omega}t)$.
The angular velocity
$\tilde{\omega} = \sqrt{4f/\rho a s_{o}}$ is small compared with
the values $10^{4-5}$ rad s$^{-1}$ for which there is efficient
coupling of shear and Alfv\'{e}n waves. ( It is $\tilde{\omega}
= 6.3\times 10^{2}$ rad s$^{-1}$ for the maximum shear
stress $10^{27}$ dyne cm$^{-2}$, $\rho= 10^{14}$ g cm$^{-3}$, and
$s_{o}=a=10^{4}$ cm.)  In principle,
adiabatic heating on the fault surface could lead to a localized
solid-liquid phase 
transition, the reduction in viscosity by many orders of magnitude
giving slip velocities much larger than $\tilde{\omega}s_{p}$.
An order of magnitude estimate of the condition that slip and
adiabatic heating should increase 
the temperature to $T_{m}$ on the fault surface is
given by
$f_{o}s_{p}\approx CT_{m}\sqrt{\left(
\kappa/\tilde{\omega}C\right)}$,
where $\kappa$ is the thermal conductivity and $C$ the specific heat. At
a typical density of $8\times 10^{13}$ g cm$^{-3}$, $T_{m} = 5.8\times
10^{9}$ K for the lattice parameters given by Negele \& Vautherin (1973).
The specific heat is almost entirely that of normal neutrons;
$C= 4.9\times 10^{20}$ erg cm$^{-3}$K$^{-1}$ for an effective mass of
$0.8 m_{n}$ (see Pines \& Nozieres 1966).  The thermal conductivity
$\kappa = 1 \times 10^{20}$
erg cm$^{-1}$s$^{-1}$K$^{-1}$ has been estimated from the electron
conductivity results given by Gnedin et al (2001).
For $f_{o}\approx 10^{27}$ dyne cm$^{-2}$, this condition shows that
a local slip distance exceeding $s_{p}\approx 50$ cm would be required
to increase
the fault surface temperature to $T_{m}$.  Thus a slip of this order
of magnitude could produce melting and subsequent
slip velocities much larger than $\tilde{\omega}s_{p}$.  But the
assumption of a single shear band in the form of a cylindrical fault
surface is unique and
does not appear realistic.  The general case is of more complex
movements, with considerable deformation, in the form of
plastic flow where
stress transfer produces many distinct shear bands with small
slip distances not satisfying the melting condition. 
In such flow, the velocity $v \approx c_{s}\Delta\epsilon$
and acceleration $\dot{v}\approx\omega v$ necessary for efficient shear
wave generation at angular velocity $\omega$ are not reached.




\section{Conclusions}

Our conclusion is that the strain relaxation conditions
necessary for the generation of shear waves
are not present in neutron stars. This need not have a great
impact on theories of the pulsar glitch phenomena.  In the theory
of Ruderman, Zhu \& Chen (1998), the movements caused by Maxwell
stress and by spin-down of the neutron superfluid in the core can be
many orders of magnitude slower than those associated with brittle
fracture and yet remain consistent with the observed upper limit
on the spin-up time for Vela glitches.  But the hypothesis that 
a plastic-brittle transition, on cooling to internal temperature
$T \approx 10^{-1}T_{m}$ at ages between $10^{3}$ and $10^{4}$ yr,
is responsible for the absence of large glitches in very young 
pulsars is not consistent with the properties of neutron-star
matter fault-planes deduced here. These may also be useful in determining
the mechanism for energy release in AXP and SGR sources (Thompson \&
Duncan 1996, Thompson et al 2000). These authors assume that a
plastic-brittle transition occurs at
internal magnetic flux densities near to $B_{\mu} = \sqrt{4\pi\mu}$,
and that brittle fractures at $B < B_{\mu}$
are the source of Alfv\'{e}n wave generation (Blaes et al 1989)
leading to both  quiescent and burst emission.  But the
mode of fracture is determined by the nature of the solid,
and its occurrence depends not on the
absolute size of
Maxwell stress components but on the extent to which they deviate
from values that would give an equilibrium in a completely liquid
star.  Our conclusion, that
brittle fracture is not possible at any temperature,
indicates that other mechanisms (see, for example,
Thompson, Lyutikov \& Kulkarni 2002; Mereghetti et al 2002) need
to be considered in more detail.

\clearpage


\begin{figure}
\includegraphics{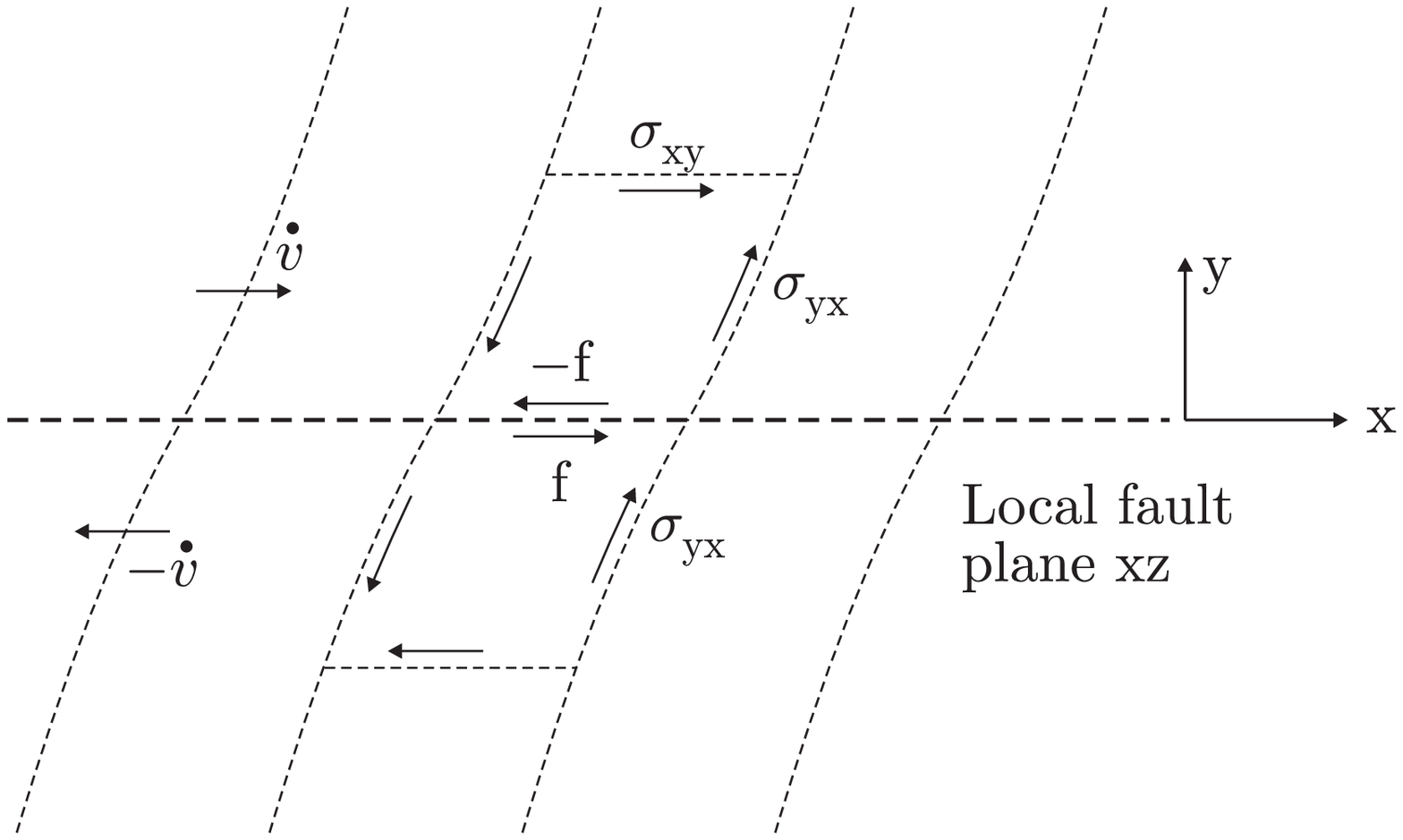}
\caption{For in-plane shear, with displacement shown by a series of
constant surfaces in Lagrangian coordinates, the components of the stress
tensor acting on volume elements on opposite sides of the $y=0$ fault-plane
are $f=\sigma_{xy}$ before failure. In ideal brittle failure,
$f=0$ immediately afterwards and strain relaxation occurs with
acceleration $\dot{v}$ such that stress energy is efficiently converted
to shear waves.  For neutron-star matter, $f$ can exhibit no
sudden decrease. \label{fig1}}
\end{figure}

\end{document}